\documentclass{elsart}
\usepackage{graphicx}
\begin{document}
\raggedbottom
\hspace*{\fill}SMU-HEP 00-02\linebreak
\begin{frontmatter}
\title{The CLEO~III Upgrade} 
\author{Thomas Coan\thanksref{doe1}} 

\address{Physics
Department, Southern Methodist University, Dallas, TX 75275, USA}

\thanks[doe1]{Supported by the U.S. Department of Energy}
\begin{abstract}

The CLEO detector at the Cornell Electron Storage Ring (CESR) is
completing a major upgrade that significantly extends its physics
reach in the study of heavy quarks and leptons.  The major elements of
this detector upgrade are the installation of a ``fast'' ring imaging
Cherenkov detector, a new central drift chamber and a new barrel-style
silicon strip vertex detector. Trigger and data acquisition systems
are also upgraded. We summarize the status of these detector
subsystems as well as the parallel CESR upgrade.

\end{abstract}
\end{frontmatter}
\normalsize

\vskip 3.cm
{\em Invited talk at ``The Seventh International Conference on 
Instrumenation for Colliding Beam Physics,'' Hamamatsu, Japan,
Nov. 15-19, 1999.}
\newpage

\section{Introduction}

The CLEO physics program is a broad effort that emphasizes the study
of $b$ and $c$ quark decays, $\tau$ lepton decays and two photon
mediated reactions.  Data are collected at or
slightly below the $\Upsilon(4S)$ resonance ($E\simeq 5.3\,$GeV)
produced by $e^+e^-$ collisions at the symmetric, one ring Cornell
Electron Storage Ring (CESR).

Important elements of the CLEO physics program include studies of rare
B decays, determination of CKM matrix elements, and possible
observation of time-independent CP violation. Examples include a
measurement of the ratio of CKM matrix elements $V_{td}/V_{ts}$ by
measuring $\Gamma(B\rightarrow \rho\gamma)/\Gamma(B\rightarrow
K^{\star}\gamma)$, a measurement of $V_{ub}$ by measuring
$\Gamma(B\rightarrow \rho(\omega)l\nu)$, and possible observation of
$CP$ violation in $B\rightarrow K\pi$ decays. These
examples, and others, involve rare reactions whose final states
include pions and kaons. Large integrated luminosity and efficient
charged hadron identification are necessary to reduce
backgrounds, as well as statistical and systematic errors.

The CLEO~III upgrade comprises two major efforts: an enhanced CLEO
detector and an upgraded CESR. To substantially enlarge data sets,
CESR is modified to boost its instantaneous luminosity to
{$L\sim2\times10^{33}\,{\rm cm^{-2}sec^{-1}}$}.  To utilize the boost
in integrated luminosity, key CLEO sub-detectors have been designed
and installed to reduce physics backgrounds present in many CLEO
analyses. CLEO~III contains a new silicon tracking device, a new drift
chamber and a novel dedicated particle identification device which
replaces a time-of-flight counter.  The trigger and data acquisition
systems are upgraded to accommodate the higher data taking rate. There
are minor modifications to the rest of the CLEO sub-detectors.

\section{CESR Improvements}

To increase the threshold for longitudinal beam instability, a past
limitation to higher luminosity, and to provide adequate power to the
larger amperage beams, CESR has replaced completely all of its room
temperature 5-cell copper cavities in its RF system by 4 single cell
superconducting cavities. These new cavities are designed to produce
an accelerating gradient of $10\,$MV/m, allowing the bunch length in
CESR to be reduced to $13\,$mm, to transmit $325\,$kW to the beam, and
to raise the threshold for longitudinal instability above $1\,$A total
beam current.

Installation of new interaction region (IR) optics is scheduled for
the Spring of 2000 to reduce the amplitude of the vertical betatron
function $\beta$ at the interaction point (IP) to
$\beta^{\star}_v=13\,$mm.  These optics are two pairs of a set of
magnets consisting of a permanent quadrupole magnet and two
superconducting (SC) quadrupoles. Each permanent neodymium iron boron
magnet provides vertical focusing and is positioned just $337\,$mm
from the IP.  Most of the focusing is provided by the four identical
SC magnets that can produce gradients up to $48.4\,$T/m at $1225\,$A.
The entire set of IR optics is contained within $\pm2.55\,$m of the
IP.  The proximity of the IR optics to the IP necessitates the
installation of a new tracking system for CLEO since these new magnets
intrude into the volume of CLEO's former configuration.

After the upgrade, CESR will collide beams of 9 nine trains each, with
each train composed of 5 bunches separated by $14\,$ns. The peak total
beam current will be $1\,$A. Opposing beams will continue to circulate
in a pretzel orbit scheme and will collide at the IP with a small
horizontal crossing angle $\theta_C=2.7\,$mrad. The expected
luminosity is $L\sim2\times10^{33}\,{\rm cm^{-2}sec^{-1}}$.

\section{Particle Identification}

\begin{figure}[htbp]
\centering

\includegraphics[scale=0.55]{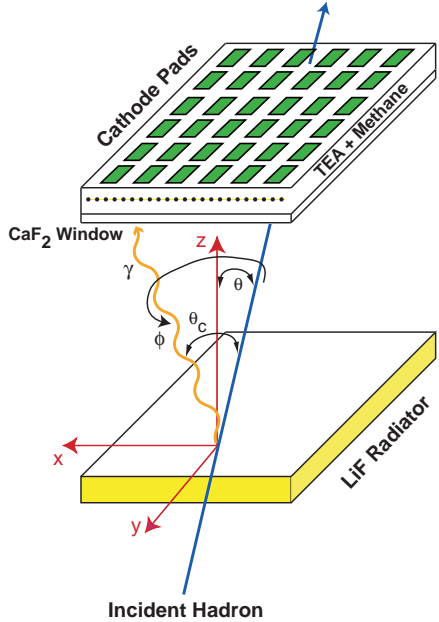}
\caption{\label{fig1}{The CLEO III RICH scheme. See text for explanation.}}
\end{figure}

\begin{figure}[htbp]
\centering

\includegraphics[scale=0.50]{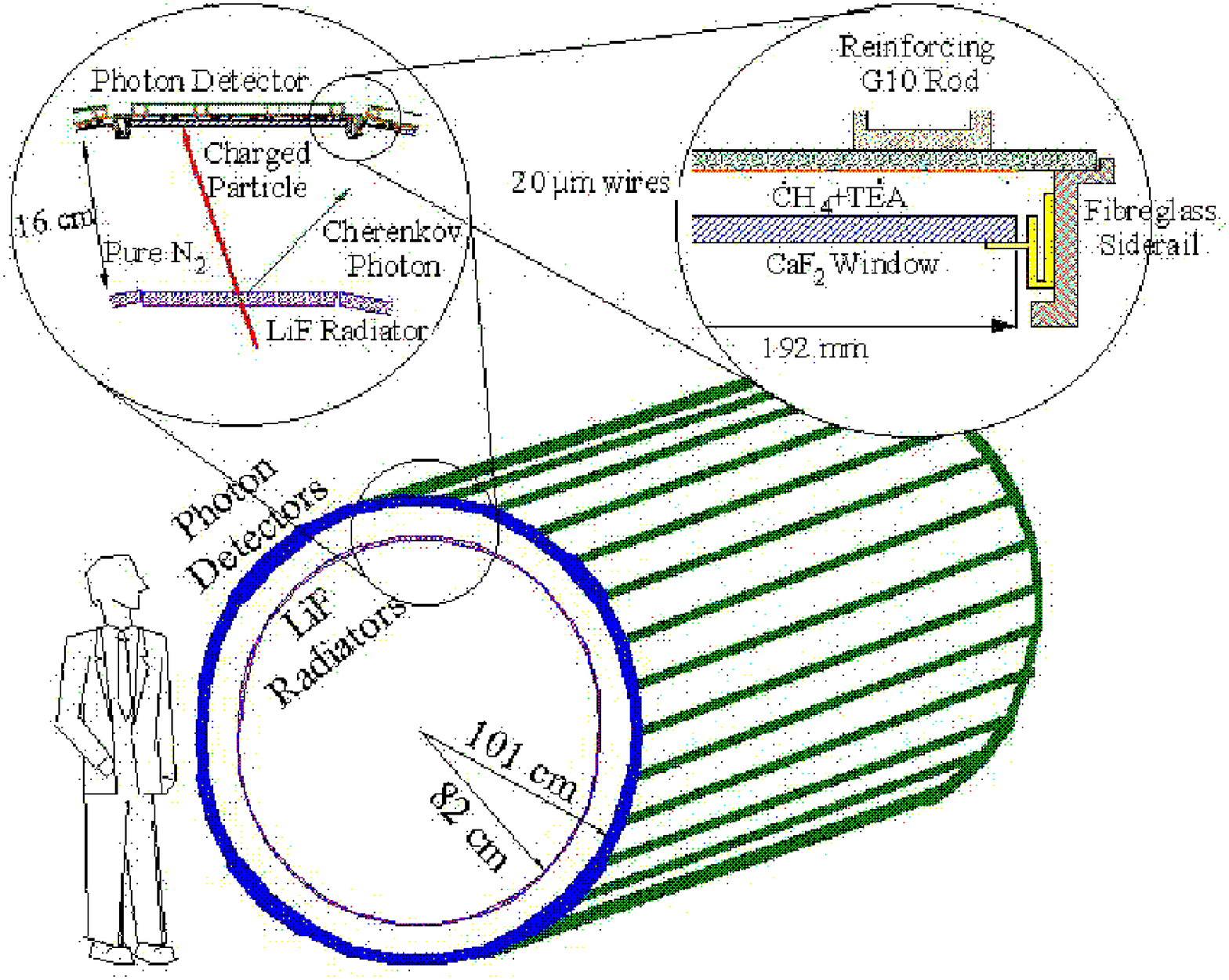}
\caption{\label{fig2}{The key elements of the CLEO~III RICH detector. A
quasi-cylinder of LiF radiator tiles is nested inside another one,
with identical azimuthal symmetry, comprised of MWPCs.  Detail of an
MWPC is shown in the top half of the figure.}}
\end{figure}

A fast ring imaging Cherenkov detector (RICH) is CLEO's technology
choice for particle identification (PID).  The general scheme is shown
in figure~\ref{fig1} where a charged hadron exits the IP, traverses a solid LiF
radiator and produces Cherenkov photons in the familiar cone
pattern. The Cherenkov cone expands in an uninstrumented volume
outside the radiator before being intercepted by a thin multiwire
proportional chamber (MWPC) filled with a photon conversion gas of
triethylamine (TEA) and methane.  The photoelectrons are amplified
near the MWPC anode wires and the resultant charge is then
capacitively coupled to cathode pads at the rear of the MWPC which are
then read out by front end electronics.  The Cherenkov angle is
inferred from the pattern of hit cathode pads. The actual
implementation of this scheme is shown in figure~\ref{fig2} where a
quasi-cylinder of lithium fluoride (LiF) is surrounded at slightly
larger radius by a quasi-cylinder of multi-wire proportional counters
(MWPCs) with the same azimuthal symmetry. The device covers 80\% of
the solid angle.

The design goal for the RICH detector alone is to provide at least
$3\,\sigma$ $\pi/K$ Cherenkov angle separation at momentum
$p=2.8\,$GeV/c, the maximum possible for $B$ decay daughters at
CLEO. Combining a RICH measurement with a current $2\,\sigma$ $\d E/\d
x$ measurement from the drift chamber for particle momentum exceeding
$p=2.2\,$GeV/c, the overall CLEO design goal is to provide $>
3.6\,\sigma$ $\pi/K$ separation at high momentum.  This corresponds to
a Cherenkov angle resolution of 4~mrad per charged track, equivalent
to an angular resolution of 14~mrad/photoelectron and 12
photoelectrons per track.

The radiator material is lithium fluoride (LiF) in the form of tiles
with planar dimensions $17\,{\rm cm} \times 17.5\,{\rm cm}$ and mean
thickness $10\,$mm. The tiles are arrayed in a regular pattern to form
a quasi-cylinder of $1.6\,$m diameter and 30-fold azimuthal
symmetry. LiF is selected to accommodate the narrow wavelength
$\lambda$ region $(135\,{\rm nm} < \lambda < 165\,{\rm nm})$ where the
TEA-methane quantum efficiency is usable, to minimize the total
radiation thickness in front of CLEO's calorimeter, and to minimize
the chromatic variation in the radiator index of refraction
$n(\lambda)$, which directly affects the Cherenkov angle resolution.

The detected Cherenkov photon patterns form images in the MWPCs of
conic sections distorted by refractive effects. Roughly half the
Cherenkov cone is trapped inside a radiator crystal by total internal
reflection (TIR). Indeed, if all radiator crystals were
parallelepipeds, then charged tracks traveling at angle
$\theta=90^{\circ}$ with respect to the beamline would have {\it
all\/} of their Cherenkov photons trapped inside a radiator crystal by
TIR. To prevent this, all radiator crystals within
$\theta=90^{\circ}\pm22^{\circ}$ of the beamline have their top
surface cut in a sawtooth pattern\cite{saws} so that Cherenkov photons
can strike the upper crystal surface at incidence angles below the
critical value for TIR.

A cross-sectional view of one of the 30 MWPCs is shown in
the upper portion of figure~2. Each MWPC has a $250 \times 20\,{\rm cm^2}$
rectangular footprint with 70 Au-plated tungsten anode wires of
$20\,\mu$m diameter running longitudinally. Its $2\,$mm thick front
window is built from 8 rectangular ($30\times 19\, {\rm cm^2}$) $\rm
CaF_2$ crystals that have $100\,\mu$m Ag silver traces applied to
them. The MWPC cathode is a printed circuit board with $7.5\times 8\,
{\rm mm^2}$ Au-plated Cu pads. The MWPC is run at a typical gain
$g\sim 4\times 10^4$.

The RICH front~end readout electronics is based on a dedicated
64-channel VLSI chip comprised of a preamplifier, a shaper with a
programmable peaking time, sample and hold circuitry, and an analog
output multiplexer.  Through a via, each of the total 230,400 MWPC
pads is connected to a preamplifier input channel.  The low noise
front end has a linear response for input signals up to $\sim 3 \times
10^5$~electrons with a measured equivalent noise charge ENC, for
detector capacitance $C$, ${\rm ENC} = 130 e^- + (9e^-/{\rm pF)}C\sim
150\,e^-$. The front end output is a differential current transmitted
serially to off-detector data acquisition boards.

A test beam \cite{tbeam} performed with 100~Gev muons incident on
planar and sawtooth radiator crystals imaged by 2 MWPCs yields, for
the sawtooth crystal, a mean number of photoelectrons per track
$N_{PE}=13.5$, a Cherenkov angle resolution per photoelectron
$\sigma_{PE}=10.2\,$mrad, and an overall Cherenkov angle resolution
per track $\sigma_{TRK}=4.5\,$mrad. Similar results are obtained for
the planar crystal. For the CLEO detector itself, with has improved
tracking and larger acceptance, we expect $\sigma_{TRK}=2.9\,$mrad,
satisfying our design goal.

\section{Central Drift Chamber}

The CLEO~III central wire tracker is a 47 layer drift chamber with
wire layers extending radially from $132\,$mm to $790\,$mm.  Outer
cathode strips at $797\,$mm radius provide a final z-position
measurement.  The first 16 wire layers are axial and the remaining 31
layers are small angle ($\simeq 25\,$mrad) stereo. Tungsten sense
wires are $20\,\mu$m diameter and the aluminum field wires are
$110\,\mu$m diameter. With a 3:1 field:sense wire ratio, the 9796
drift cells are nearly square with $\simeq14\,$mm width. The average
chamber radiation thickness $X/X_0=3.3\%$ for normally incident
particles. The drift gas is a 60:40 mixture of helium-propane,
selected for its long radiation length ($X_0 > 569\,$m), and its
favorable drift and ionization properties.

\begin{figure}[htbp]
\centering

\includegraphics[scale=0.50]{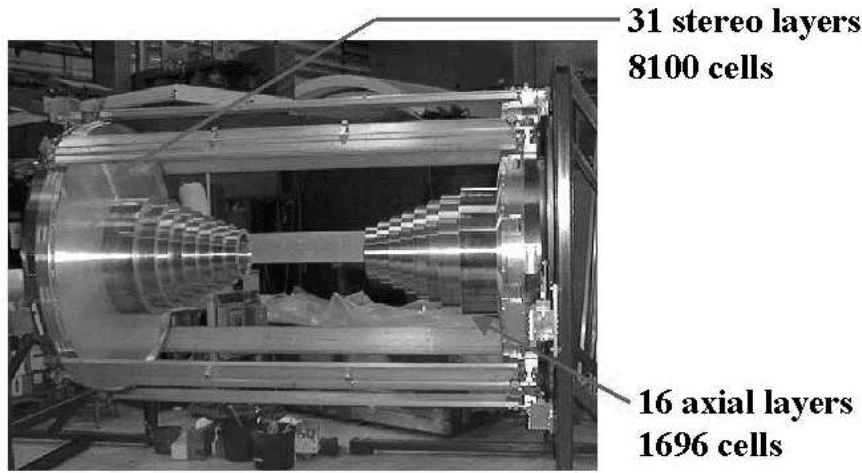}

\caption{\label{fig3}{The CLEO III drift chamber under construction. Note the
stepped portion of the endplate used to accommodate the CESR final
focus quadrupoles. The number of drift cells is also indicated.}}
\end{figure}

The mechanical structure of the chamber accommodates the close
positioning of the final focus quadrupoles by having a stepped
endplate to hold and position the axial wire layers.  See
figure~\ref{fig3}.  A conical endplate holds and positions the
stereo wires. The wire tension is resisted solely by outer skins that
hold the cathode strips. An inner tube provides a gas seal only.

Sense wire signals are read by preamplifiers that drive differential
analog current signals to special purpose ``time-charge-trigger'' (TQT)
front end boards that provide signals for timing, charge measurement
and the trigger. Signals received by the TQT boards are split into
parallel timing/charge and trigger paths. The boards have independent
thresholds for the discriminated timing and trigger signals to permit
a low trigger rate without degrading timing resolution. The output of
the timing and charge-to-time circuits are multiplexed onto the same
channel of a LeCroy 1877S Fastbus multi-hit TDC. The discriminated
trigger signal is sent to the trigger system.

\section{Silicon Vertex Detector}

CLEO III includes a 4-layer barrel-style silicon strip vertex detector
with double-sided sensor layers located at radii between $2.5\,$cm and
$10.1\,$cm and covering 93\% of the solid angle. To simplify overall
ladder production, all 447 of the $300\,\mu$m thick sensors are
identical and contain no coupling capacitors or bias resistors.
Ladder lengths span $16\,$cm at the inner radius to $53.3\,$cm at the
outer radius. The n-strips measure the $r-\phi$ coordinate and have a
$50\,\mu$m pitch, while the p-strips measure the z-coordinate and have
a $100\,\mu$m pitch. Considerable care has been taken to minimize
sensor capacitance\cite{shipsey}.  Approximately $125,000$ total
strips are read out.

The vertex detector mechanical design minimizes material in the
fiducial region by placing the detector's major mechanical support
structure and readout electronics outside this volume.  Ladders are
mechanically stiffened by gluing V-shaped CVD diamond beams to them
and are supported at their ends by copper end cones through kinematic
mounts.  A flexible kapton circuit electrically connects the sensors
to the front end electronics mounted on BeO hybrids that are attached
directly to the copper cones for efficient cooling. A thin carbon
fiber tube joins the opposite copper end cones.

A signal-to-noise ratio $S/N>15$:1 for all sensor layers is the
overall detector readout goal. The front end electronics is built from
3 different chips mounted on double-side hybrids.  For detector
biasing and decoupling, a special 128 channel $R/C$ chip is used with
$R=150\,{\rm M}\Omega$ and $C=150\,$pF.  Sensor signals are amplified
and shaped by a 128 channel multi-stage
preamplifier/shaper\cite{kagan} with a $125\,$fF feedback capacitor,
200 mV/mip gain, and a $2\,\mu$s shaping time. The circuit includes a
variable gain stage to accommodate the range in ladder capacitance due
to different ladder lengths. The measured equivalent noise charge ENC
for the preamplifier/shaper is, for sensor capacitance $C$, $ENC \sim
150\,e^- + (5.5e^-/{\rm pF})C$. The final front end chip is a 128
channel ``SVX\_CLEO'' digitizer/sparsifier based on the FNAL/LBL
SVX\_II(b)\cite{milgromme}. This chip incorporates a comparator and an
8 bit ADC for each channel. The digitized results can then be
sparsified using a chip-wide threshold before being readout serially
by an off-detector VME-based data board.

\section{Trigger System}

The CLEO~III trigger system has a single hardware level with a
$2\,\mu$s trigger decision time. Since this is much longer than the
$14\,$ns bunch-to-bunch-spacing, a trigger pipeline is used with all
trigger processing done in $42\,$ns steps. The level 1 trigger
information originates from just 2 detector subsystems: the central
drift chamber and the calorimeter.  The tracking trigger is built from
an ``axial'' trigger, derived from hits on the 16 axial layers of the
drift chamber, and a ``stereo'' trigger, derived from hits on the
chamber's 31 stereo layers. The calorimeter trigger is derived from
energy deposition in the 7800 thalium doped CsI calorimeter crystals.

The relatively small number of axial wires (1696) permits the full set
to be examined for all possible hit patterns caused by tracks with
transverse momentum $p_{\perp}> 200\,$MeV/c. Up to two hits from a
valid track can be missing from each of the inner and outer set of 8
wires. Hit patterns are compared against look-up tables for valid
tracks using field programmable gate arrays (FPGAs) and up to two hits
are allowed to be missing from each of the inner and outer set of 8
wires.  Additional circuitry generates track count and topological
information to correlate axial and stereo tracks.

The relatively large number of stereo wires (8384) does not make the
wire-by-wire hit examination used for the axial layers practical.
Instead, stereo wires are grouped into 4x4 arrays and then patterns of
hit arrays are compared against those for valid tracks via FPGA lookup
logic. Not all arrays along an arc need be hit to satisfy the logic
which accommodates tracks with transverse momentum
$p_{\perp}>250\,$MeV/c.

The calorimeter trigger has been substantially upgraded to increase
its efficiency, particularly at low energies. Previously, the light
output from 16 crystals was summed and shaped, and then compared
against low- and high-level energy thresholds. However, energy
deposited in the calorimeter associated with a track is typically
shared among several crystals.  If these crystals lie on the border
between adjacent groups of 16 crystals, then it is possible for the
separate summed signals to fall below one or both of the
thresholds. To eliminate this effect, additional `tiling' circuitry is
used to sum sets of 16 crystals into $2\times 2$ arrays or `tiles' of
64 crystals who total signal is then compared against three
thresholds. The tiles overlap so that the problem of energy sharing
between crystals is eliminated.  Other circuitry is used to properly
rescale, if necessary, the apparent transverse shower size produced by
the tiling circuitry.

\section{Conclusion}
The superconducting RF cavities, the RICH detector and the central
drift chamber are all installed, and CESR and CLEO are now conducting an
engineering run scheduled from November 1999 to February 2000.  The
major sub-detector not quite complete is the silicon vertex detector,
due for installation in February 2000, after which physics running
commences. The final focus quadrupoles for CESR are due in Spring 2000
and their installation is the final element of the complete CLEO/CESR
upgrade.

\section{ Acknowledgments}
The generous assistance of K. Ecklund, V. Fadeyev, R. Mountain and
M. Selen in preparing this talk is appreciated.

\vfill
\end{document}